\documentclass[prd,preprintnumbers,twocolumn,eqsecnum,floatfix,a4paper,nofootinbib,superscriptaddress]{revtex4}
\usepackage{color}
\usepackage{calc}
\usepackage{amsmath,amssymb,graphicx}
\usepackage{amssymb,amsmath}
\usepackage{bm}
\usepackage{microtype}
\usepackage{booktabs}
\usepackage{times}
\usepackage{subfigure}
\usepackage[varg]{txfonts}
\usepackage[colorlinks, pdfborder={0 0 0}]{hyperref}
\usepackage[utf8]{inputenc}
\definecolor{LinkColor}{rgb}{0.75, 0, 0}
\definecolor{CiteColor}{rgb}{0, 0.5, 0.5}
\definecolor{UrlColor}{rgb}{0, 0, 0.75}
\hypersetup{linkcolor=LinkColor}
\hypersetup{citecolor=CiteColor}
\hypersetup{urlcolor=UrlColor}
\allowdisplaybreaks
\DeclareFontFamily{OT1}{pzc}{}
\DeclareFontShape{OT1}{pzc}{m}{it}{<-> s * [1.10] pzcmi7t}{}
\DeclareMathAlphabet{\mathpzc}{OT1}{pzc}{m}{it}

\newcommand{\h}{\mathpzc{h}}

\newcommand{\hlm}{\mathpzc{h}_{\ell m}}

\newcommand{\Ylm}{{Y}^{-2}_{\ell m}}

\newcommand{\blambda}{\bm{\lambda}}
\newcommand{\btheta}{\bm{\theta}}
\newcommand{\bxi}{\bm{\xi}}

\newcommand{\n}{\mathbf{n}}

\begin{document}

\newcommand{\be}{\begin{equation}}
\newcommand{\ee}{\end{equation}}
\newcommand{\ber}{\begin{eqnarray}}
\newcommand{\eer}{\end{eqnarray}}
\def\bea{\begin{eqnarray}}
\def\eea{\end{eqnarray}}
\newcommand{\etal}{\emph{et al}}

\title{Testing the ``no-hair'' nature of binary black holes \\using the consistency of multipolar gravitational radiation}
\author{Tousif Islam}
\affiliation{International Centre for Theoretical Sciences, Tata Institute of Fundamental Research, Bangalore 560089, India}
\affiliation{Center for Scientific Computation and Visualization Research, University of Massachusetts Dartmouth, Dartmouth, MA-02740, USA}
\author{Ajit Kumar Mehta}
\affiliation{International Centre for Theoretical Sciences, Tata Institute of Fundamental Research, Bangalore 560089, India}
\author{Abhirup Ghosh}
\affiliation{Max Planck Institute for Gravitational Physics (Albert Einstein Institute), D-14476 Potsdam-Golm, Germany}
\affiliation{International Centre for Theoretical Sciences, Tata Institute of Fundamental Research, Bangalore 560089, India}
\author{Vijay Varma}
\affiliation{Theoretical Astrophysics, 350-17, California Institute of Technology, Pasadena, CA 91125, USA}
\author{Parameswaran~Ajith}
\affiliation{International Centre for Theoretical Sciences, Tata Institute of Fundamental Research, Bangalore 560089, India}
\affiliation{Canadian Institute for Advanced Research, CIFAR Azrieli Global Scholar, MaRS Centre, West Tower, 661 University Ave., Suite 505, Toronto, ON M5G 1M1, Canada}
\author{B.~S.~Sathyaprakash}
\affiliation{Department of Physics and Department of Astronomy and Astrophysics, The Pennsylvania State University, University Park, PA 16802, USA}
\affiliation{School of Physics and Astronomy, Cardiff University, Cardiff, CF24 3AA, UK}

\begin{abstract}
Gravitational-wave (GW) observations of binary black holes offer the best probes of the relativistic, strong-field regime of gravity. Gravitational radiation, in the leading order is quadrupolar. However, non-quadrupole (higher order) modes make appreciable contribution to the radiation from binary black holes with large mass ratios and misaligned spins. The multipolar structure of the radiation is fully determined by the intrinsic parameters (masses and spin angular momenta of the companion black holes) of a binary in quasi-circular orbit. Following our previous work~\cite{Dhanpal:2018ufk}, we develop multiple ways of testing the consistency of the observed GW signal with the expected multipolar structure of radiation from binary black holes in general relativity. We call this a ``no-hair'' test of binary black holes as this is similar to testing the ``no-hair'' theorem for isolated black holes through mutual consistency of the quasi-normal mode spectrum. We use Bayesian inference to on simulated GW signals that are consistent/inconsistent with binary black holes in GR to demonstrate the power of the proposed tests. We also make estimate systematic errors arising as a result of neglecting companion spins. 
\end{abstract}
\preprint{}
\maketitle
\section{Introduction}

Recent gravitational-wave (GW) observations of coalescing compact binaries by LIGO and Virgo have provided a unique test bed for gravity ~\cite{LSC_2016grtests,gw170104,ligo2019tests,abbott2019tests,abbott2017gravitational}. Due to their high compactness, black holes and neutron stars in coalescing binaries are able to approach each other in close separations (comparable to their gravitational radii)~\cite{GW150914}. They also move with speeds close to the speed of light before they merge. As a result the final orbits of their inspiral and the subsequent merger will probe the relativistic strong-field regime. The subsequent formation of a nascent black hole also offers interesting tests of the nature of the black hole through the study of its perturbations~\cite{Berti:2009kk}. In addition, the GW observations also allow us to study various possible propagation effects of GWs~\cite{samajdar2017projected}, including dispersion~\cite{Will:1997bb} and damping, and to constrain the presence of additional polarization modes that are absent in general relativity (GR)~\cite{isi2017probing}. In addition, multi-messenger observations of a compact binary merger allow us to measure the speed of GWs as well as to constrain violations of equivalence principle, Lorentz invariance violations and the presence of extra dimensions~\cite{abbott2019tests,abbott2017gravitational,Pardo:2018ipy}.

One of the powerful probes of the nature of black holes that can be performed using GW observations is to test the ``no-hair'' theorem in GR --- the prediction that a stationary black hole in GR can be fully described solely by its mass, spin angular momentum and electric charge~\cite{Israel:1967,Israel:1968,Carter:1978}. As a consequence of this, the frequencies of the {quasi-normal modes}~\cite{Vishveshwara:1970zz,Press:1971wr,Chandrasekhar:1975zza} of the gravitational radiation from a perturbed black hole are fully determined by these parameters. If we are able to measure three quasi-normal mode frequencies, this allows, in principle, the determination of the mass, spin and charge of the black hole. Since astrophysical black holes are unlikely to possess significant electric charge, a black hole's mass and spin can be determined from the measurement of just two quasi-normal mode frequencies. If we are able to measure more than two quasi-normal modes, the black hole mass and spin estimated from multiple modes have to consistent with each other; otherwise it will point to a violation of the no-hair theorem~\cite{Dreyer:2003bv}.

In a similar fashion we expect the dynamics and gravitational radiation from a binary black hole system in a quasi-circular orbit to be uniquely determined by a small number of parameters (masses and spins of the black holes). Hence different multipoles (spherical harmonic modes) of the radiation have to be consistent with the same values of the black holes' masses and spins. Thus, the consistency between different modes of the observed signal is a powerful test that the radiation is produced by a binary black hole system. Inconsistency between different modes of the radiation would point to either a departure from GR, or the non-black hole nature of the compact objects. The larger signal-to-noise ratio (SNR) obtained from analyzing the full inspiral-merger-ringdown signal will give us an advantage over a the consistency test of different quasi-normal modes~\footnote{Admittedly, the test proposed in this paper is not a direct probe of the violation of the ``no-hair'' nature of isolated black holes. However, we anticipate such a violation to show up as a departure from the expected multipole structure of the binary black hole waveform.}. 


Such a no-hair test for binary black holes was presented in~\cite{Dhanpal:2018ufk}\footnote{Another test of the multipolar structure involving a parametrized phasing formula for the inspiral part of the gravitational radiation from compact binary coalescences was suggested in~\cite{kastha2018testing,kastha2019testing}.}. The main idea of this test is to test the consistency of the the source parameters estimated from the quadrupole (leading order) modes and higher order modes separately. In spirit, this idea is similar to checking the consistency of cosmological parameters estimated from the low- and high multipoles of the cosmic microwave background radiation (see, e.g.,~\cite{Aghanim:2018eyx}). In this paper we present different formulations of such a test, demonstrate their application using simulated data and present a first investigation of the systematic errors that need to be controlled before the test is applied to real GW observations.

Indeed, this test requires the higher order modes of the radiation to be detected with sufficient SNR. This entail the observation of binaries with large mass ratios and/or highly misaligned spins with high inclination angles (angle between the orbital angular momentum and the line of sight). Since GWs are primarily radiated in a direction parallel/anti-parallel to the orbital angular momentum, GW observations have a selection bias towards binaries with small inclination angles, and hence the contribution from higher modes is likely to be small for most observed systems. However, considering Advanced LIGO and Virgo are expected to detect hundreds of binary black hole mergers in the next few years, we are likely to detect a small number of high-mass ratio binaries in inclined orbits which enables this test to be performed~\cite{Dhanpal:2018ufk}. There is already preliminary evidence of higher modes in one of the binary black hole events detected by LIGO and Virgo during their second observing run~\cite{Chatziioannou:2019dsz}. 


The rest of the paper is organized as follows: Section~\ref{sec:test} presents two different formulations of the test along with the Bayesian implementation. Section~\ref{sec:simulations} presents results from this test applied to simulated GW observations of binary black holes in GR, while Sec.~\ref{sec:simulation_nonbbh} presents results from simulated observations containing deviations from binary black holes in GR. Section~\ref{sec:waveformsyst} presents a first investigation of systematic errors due to neglecting the effect of black hole spins in the GR waveforms. Finally, Sec.~\ref{sec:conclusions} presents a summary and concluding remarks. 

\section{Testing the consistency of different multipoles of the radiation}
\label{sec:test}

\subsection{Multipolar gravitational waveforms from binary black holes}

Gravitational radiation from the coalescence of a binary black hole in GR can be written as a superposition of $-2$ spin-weighted spherical harmonics ~\cite{NewmanPenrose}:
\begin{eqnarray}
\h(t; \n, \blambda) &:=& h_+(t; \n, \blambda) - i \, h_\times(t; \n, \blambda) \\
&=& \sum _{\ell=2}^{\infty} \sum _{m=-\ell}^{\ell} {\Ylm} (\n) \, {{\hlm}(t; {\blambda})}, 
\label{eq:spherical_harmonics}
\end{eqnarray}
where $h_+$ and $h_{\times}$ are the two independent polarizations of gravitational radiation, ${\Ylm}$ spherical harmonics of weight $-2$ and $\n := \{\iota, \varphi_0\}$ the direction of radiation in the source frame. The spherical harmonic modes can be computed from the full radiation as 
\begin{equation}
\hlm(t; {\blambda}) := \int_0^{2\pi}  d\varphi_0 \int_0^{\pi} \h(t; \n, \blambda) \, {\Ylm}^\star (\n) \, \sin \iota \, d\iota, 
\label{eq:sphar_modes}
\end{equation}
where the integration is over the full sphere~\footnote{Note that while we are able to theoretically compute the spherical harmonic modes of the radiation from a binary, it is not possible to estimate the modes from the observed signal, since the observed signal $\h(t; \n, \blambda) $ is a particular linear combination of the modes. This is very different, for example, from the observation of cosmic microwave background radiation where the radiation is measured over the entire sphere and hence the radiation multipoles can be computed using a decomposition similar to Eq.\eqref{eq:sphar_modes}.}. In GR, the leading order mode is the quadrupolar ($\ell = 2, m = \pm 2$) modes. The relative contribution of the higher modes to the signal $\h(t; \n, \blambda)$ depends on the total mass $M$, mass ratio $q$, spin angular momenta ${\mathbf S}_{1,2}$ and the orientation of the binary $\n$. 

The spherical harmonic modes, ${\h}_{lm}(t; \blambda)$, are uniquely determined by the \emph{intrinsic} parameters $\blambda$ of the system, i.e., the masses and spins of the two black holes (for a quasi-circular binary). Thus, by comparing these theoretical waveforms with data (see, e.g. Sec~\ref{sec:bayesian_analsysis}), one can estimate these parameters. In this work, we model the gravitational radiation from non-spinning binary black holes using the phenomenological inspiral-merger-ringdown waveform family introduced in \cite{Mehta:2017jpq}. This waveform model includes the $(\ell = 2, m=\pm1)$, $(\ell = 3, m=\pm3)$ and $(\ell = 4, m = \pm4)$  modes of the radiation over and above the dominant $(\ell = 2, m = \pm2)$ mode. The other spherical harmonic modes neglected in this waveform model only introduce an inaccuracy (mismatch) of less than 1\% in the waveforms~\cite{Mehta:2017jpq}.

\subsection{Formulation of the test}
\label{sec:formulation}

In ~\cite{Dhanpal:2018ufk}, we presented a new test of the ``no-hair'' nature of binary black holes in GR based on the consistency of different multipoles (spherical harmonic modes) of the radiation. In spirit, this involves estimating the intrinsic parameters of the binary from different multipoles of the radiation and checking their consistency. If the parameters estimated from two different modes are inconsistent with each other, this would imply that the multipolar structure of the radiation is inconsistent with what is expected from a binary black hole system in GR. In practice, we are unable to extract the different multipoles of the radiation from the observed GW signal. Hence we introduce extra parameters in the signal model that allows discrepancies between different modes and estimate those parameters along with the standard set of parameters that describe the GW signal. If the signal is consistent with that produced by a binary black hole system in GR, the additional parameters will be consistent with zero. 

\paragraph{Formulation A:}
	Following ~\cite{Dhanpal:2018ufk}, we generalize the GR waveform model Eq.~(\ref{eq:spherical_harmonics}) by allowing inconsistencies between the intrinsic parameters estimated from the dominant mode and the higher order modes by introducing a set of deviation parameters $\Delta \blambda := \{\Delta M_c, \Delta q\}$ in the higher modes: 
	\begin{eqnarray}
	\h(t; \n, \blambda, \Delta \blambda) & = &    \sum_{m = \pm2} Y^{-2}_{2m} (\n) {\h}_{2m}(t, \blambda) \nonumber \\ 
	& + &  \sum _{\text{H.O.M}} \Ylm (\n) \hlm(t, \blambda+\Delta \blambda), 	
\label{eq:test_1}
	\end{eqnarray}
where {H.O.M} indicates sum over higher order modes (all modes other than $\ell = 2, m = \pm 2$). We then simultaneously estimate the posterior distributions of $\blambda$ and $\Delta \blambda$ along with other extrinsic parameters that describe the location and orientation of the binary (see Sec.~\ref{sec:bayesian_analsysis}). 

\paragraph{Formulation B:}
In this paper, we also investigate modifications made to the amplitude of the gravitational radiation by introducing deviations to the amplitude of non-quadrupole modes, and rewriting Eq.~(\ref{eq:spherical_harmonics}) as
	\begin{eqnarray}
	{\h}(t; \n, \blambda, c_{\ell m}) & = & \sum_{m = \pm2} Y^{-2}_{2m} (\n) {\h}_{2m}(t, \blambda)   \nonumber \\ 
	& + &  \sum _{\text{H.O.M}} (1+{c_{\ell m}}) \, \Ylm (\n) \hlm(t, \blambda) ,
	\label{eq:test_2}
	\end{eqnarray}
	where $c_{\ell m}$ is a set of deviation parameters that could be different for different higher order modes. Here we simultaneously estimate the posterior distributions of $\blambda$ and ${c_{\ell m}}$  along with other extrinsic parameters that describe the location and orientation of the binary. We consider different combinations of ${c_{\ell m}}$ (details in Sec.~\ref{sec:formulationB}). 

\subsection{Bayesian analysis}
\label{sec:bayesian_analsysis}

\begin{figure}[tb] \begin{center}
		\includegraphics[width=3.4in]{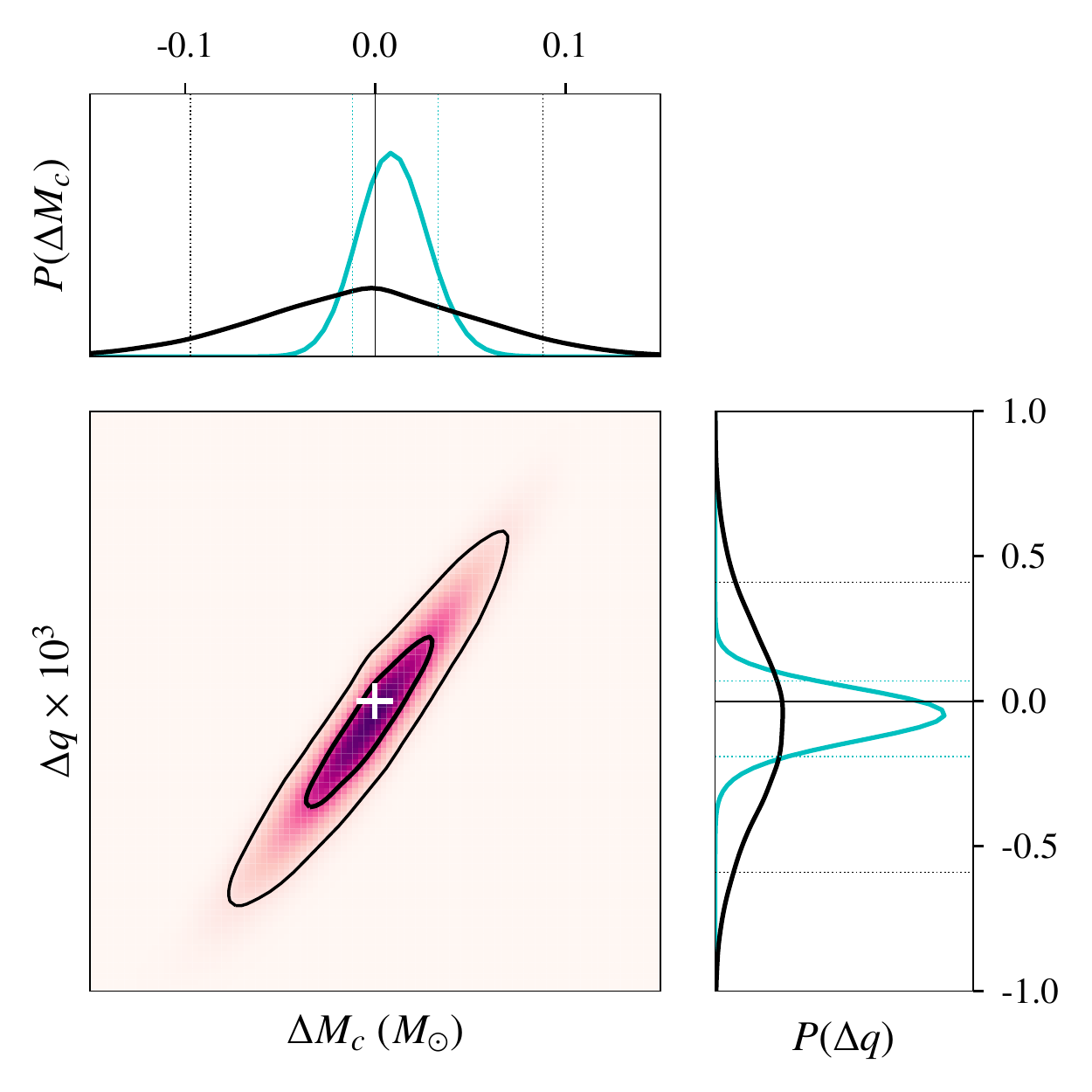}
		\caption{\emph{Middle panel}: the thick (thin) contours show the 50\% (90\%) credible regions in the joint posteriors of two parameters $\Delta M_c$ and $\Delta q$ that describe deviations in the estimated parameters using the quadrupole and non-quadrupole modes, estimated from a simulated GR signal [see Eq.~\eqref{eq:test_1} for the formulation]. \emph{Side panels}: Black histograms show the 1-dimensional posteriors in one deviation parameter (say, $\Delta M_c$) estimated from the joint posteriors, which is marginalized over the other (say, $\Delta q$). The cyan histograms show the 1-dimensional posteriors in $\Delta M_c$ and $\Delta q$ estimated from the data by introducing only one deviation parameter (say, $\Delta M_c$) at a time, keeping the other fixed (say, $\Delta q = 0$). The posteriors are fully consistent with the GR prediction of $\Delta M_c = \Delta q = 0$ (shown by a ``+'' sign in the center panel and by thin black lines in side panels). The dotted lines mark the 90\% credible regions. The simulated GR signal corresponds to a binary with total mass $M = {80}M_\odot$ and mass ratio $q = 1/9$ and an inclination angle $\iota = {60^\circ}$ observed by Advanced LIGO-Virgo detectors network with an optimal SNR of 25. SNR split in individual detectors are: 15 in LIGO-Hanford, 18.9 in LIGO-Livingston and 6.7 in Advanced Virgo.}
		\label{fig:posterior_BBH_GR_inj}
\end{center} \end{figure}

\begin{figure}[tb] \begin{center}
		\includegraphics[width=3.4in]{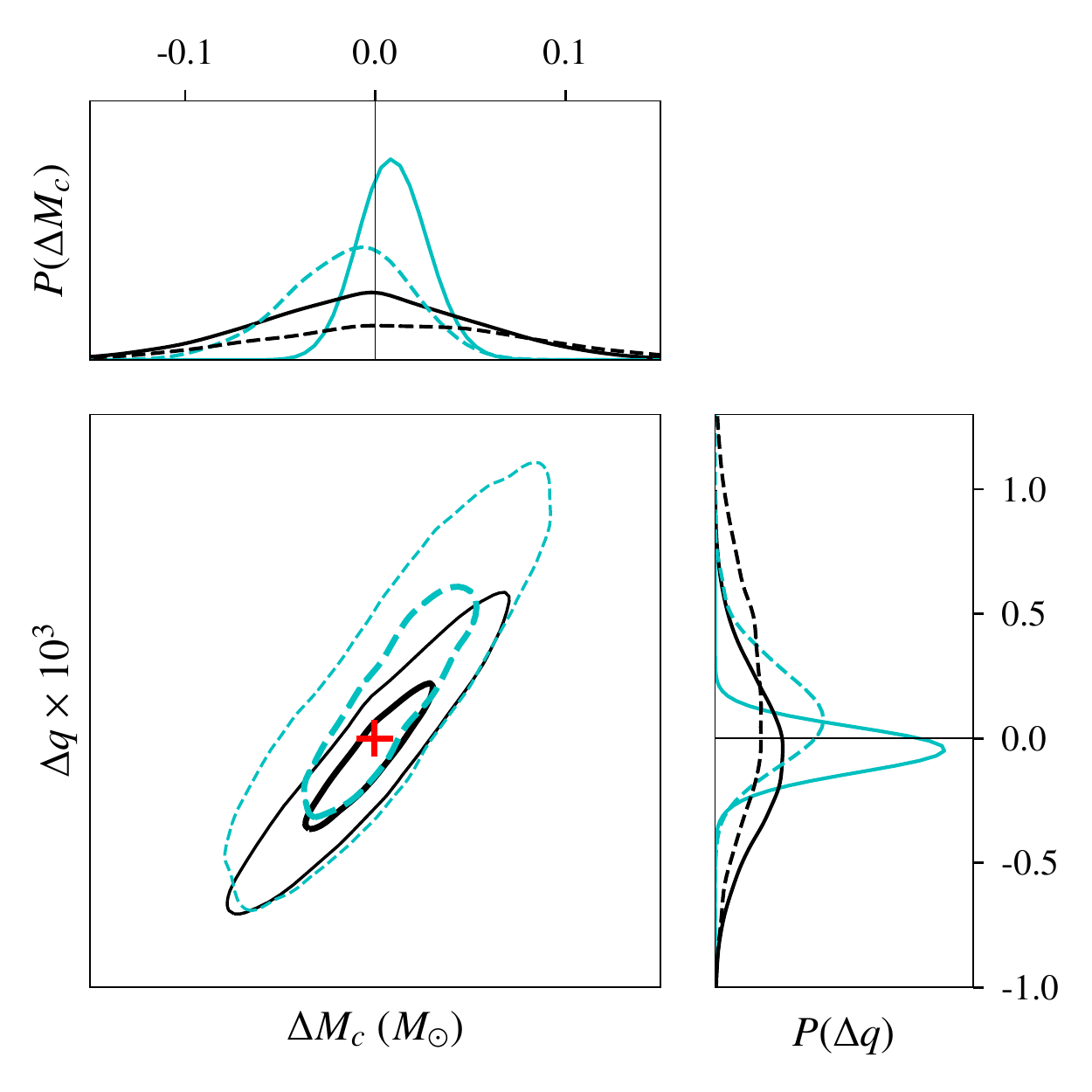}
		\caption{Comparison of the posteriors on the deviation parameters $\Delta M_c$ and $\Delta q$ estimated from a three detector observation (solid black contours; same as Fig.~\ref{fig:posterior_BBH_GR_inj}) with the same obtained using a using a single Advanced LIGO detector (dashed contours) with SNR of 25. All injection parameters are the same as the ones in Fig.~\ref{fig:posterior_BBH_GR_inj}.  It can be seen that, as expected, posteriors from the three detector observation are tighter.}
		\label{fig:hm_mcq_compare-1det_3det_GR_inj}
	\end{center} \end{figure}

 \begin{figure}[tbh]
 	\begin{center}
 		\includegraphics[scale=0.8]{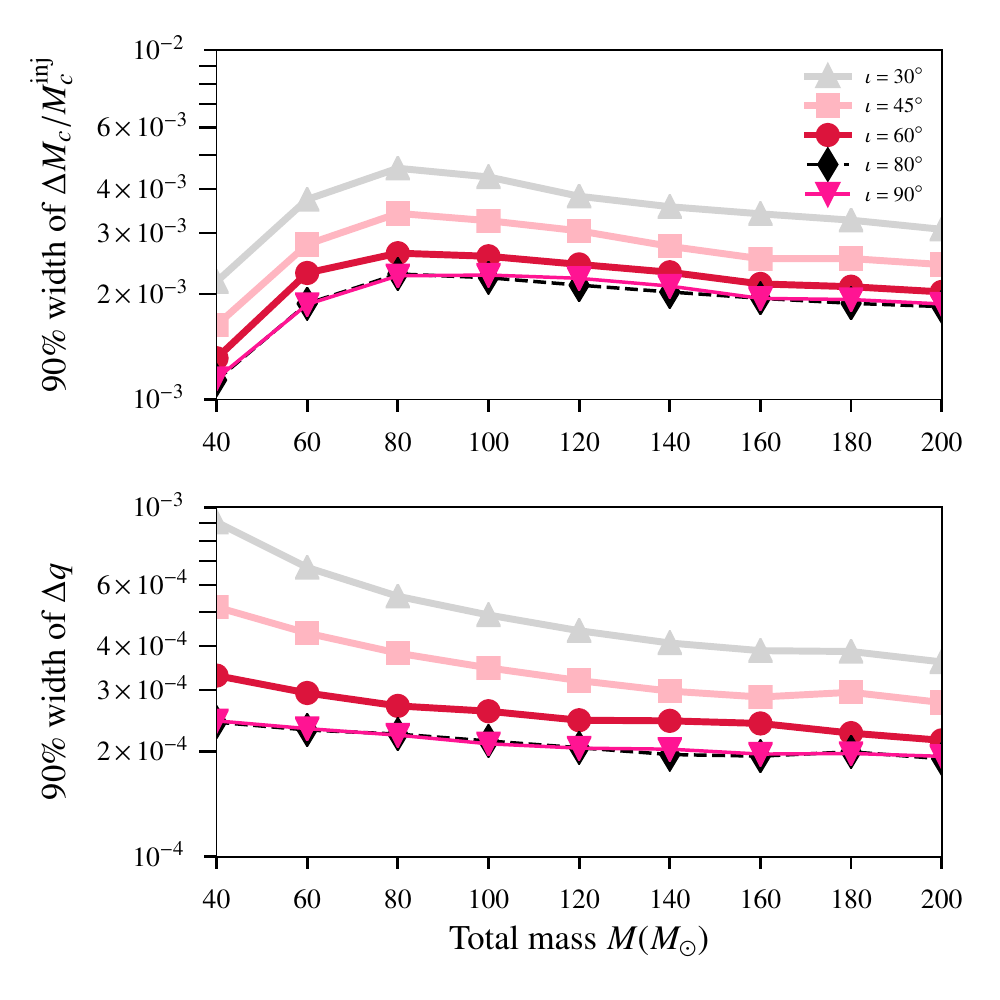}
 	\end{center} 
 	\caption{The figure shows the width of the 90$\%$ credible regions of the deviation parameters $\Delta M_c$ and $\Delta q$ for binaries with different total mass (horizontal axis) and inclination angles $\iota$ (legends). All binaries have an asymmetric mass ratio $q=1/9$.}
 	\label{fig:delmc_delq_varyingM}
 \end{figure}
 
 \begin{figure}[tbh]
 	\begin{center}
 		\includegraphics[scale=0.8]{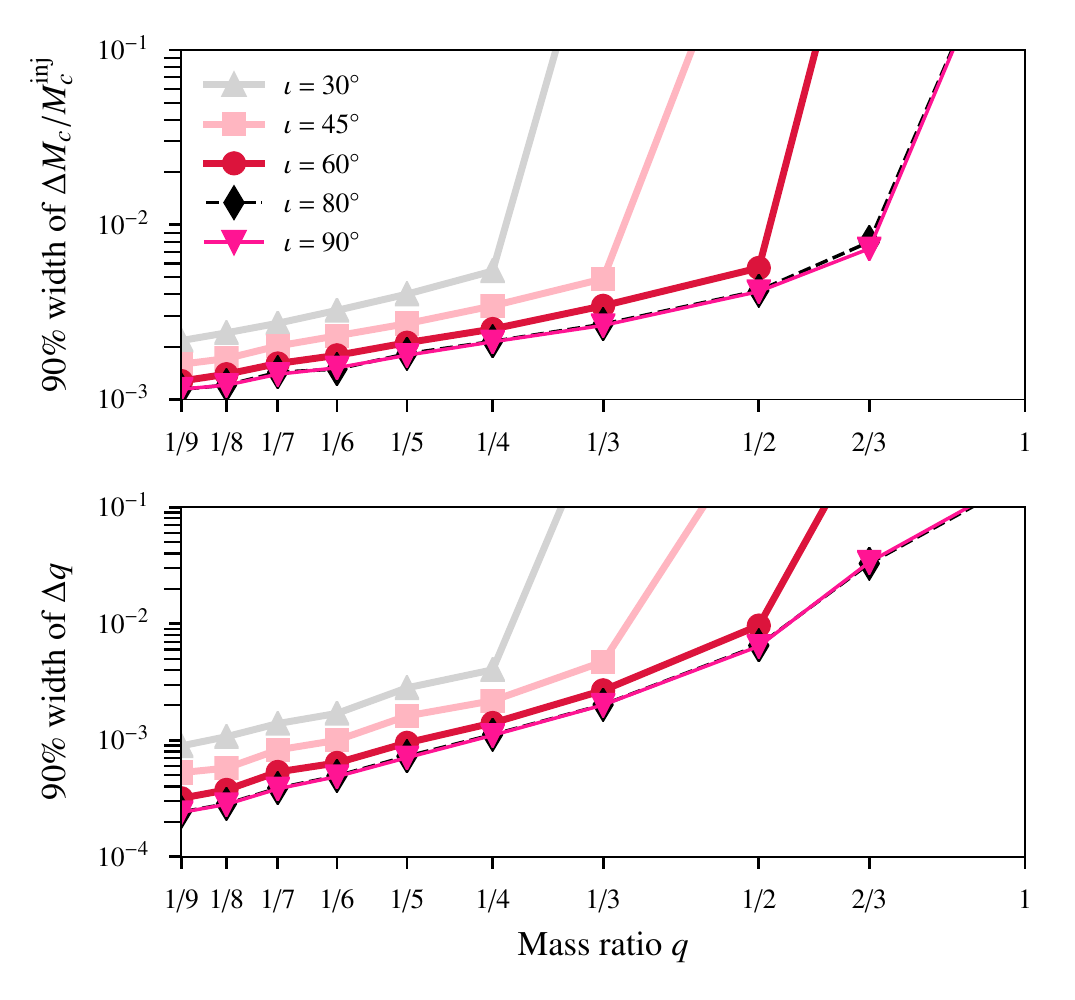}
 	\end{center} 
 	\caption{Same as Fig.~\ref{fig:delmc_delq_varyingM}, except that the horizontal axis reports the mass ratio $q$. All binaries correspond to a total mass $40M_{\odot}$.}
 	\label{fig:delmc_delq_varyingq}
 \end{figure}

Each interferometric GW detector $I$ detects a linear combination of the two polarizations $h_+$ and $h_\times$, given by 
\begin{eqnarray}
h^I(t) & = \frac{1}{d_L} \, \left [ F^I_+(\alpha, \delta, \psi) \, h_+(t-t_0; \n, \blambda) \right. \\ 
       & \left. + F^I_{\times}(\alpha, \delta, \psi)\, h_\times(t-t_0; \n, \blambda) \right]
\label{eq:det_response}
\end{eqnarray}
where ${d_L}$ is the luminosity distance to the source, $F^I_+$ and $F^I_\times$ are the antenna pattern functions of the detector $I$, $t_0$ is the time of arrival of the signal at the detector, and $(\alpha, \delta), \psi$ define the sky position and polarisation angle of the GW source, respectively. Above, we have neglected the time dependence of the antenna pattern functions, which is a good assumption for the case of the transient signals that we consider in this work. 

The noise $n(t)$ in a GW detector can be safely described, over sufficiently short time intervals, as a stationary and Gaussian random process with zero mean and a power spectral density (PSD), $S_n(f)$. In the presence of a GW signal $h(t; \btheta)$ from a binary black hole merger described by a parameter set $\btheta$ (which include the intrinsic and extrinsic parameters of the binary as well as the set of parameters describing deviations from GR), we assume that the detector data $d(t)$ is the sum of the noise and the signal, i.e.:
\begin{equation}
d(t) = n(t) + h(t; \btheta). 
\end{equation}
A (quasi-circular) non-spinning binary black hole coalescence can be completely described by a 9-dimensional parameter set $\btheta = \{ \blambda, \bxi\}$ in GR, where $\blambda = \{ M_c, q \}$ is the set of intrinsic parameters consisting of the chirp mass $M_c$ and the asymmetric mass ratio $q = m_2/m_1$ (with the convention $m_2 \leq m_1$), and $\bxi = \{ d_L, \alpha, \delta, \iota, \psi, t_0, \phi_0 \}$, is the set of extrinsic parameters consisting of the luminosity distance $d_L$, the sky position $\{\alpha, \delta\}$ and orientation $\{\psi, \iota\}$ of the binary, and the time and phase at coalescence $\{t_0, \phi_0\}$ respectively.

We use the Bayesian framework to obtain the posterior probability distribution $P({\btheta} \, | \, d)$ of the parameter set $\btheta$, through the Bayes Theorem:
\begin{equation}
P({\btheta} \, | \, d) = \frac{P({\btheta}) \, P (d \, | \, {\btheta})}{P(d)}.
\label{eq:Bayes_theorem}
\end{equation} 
where, $P({\btheta})$ denotes the prior probability distribution of the parameters, and $P (d \, | \, {\btheta})$ is the likelihood function, the probability of observing data $d(t)$ given the model parameters $\btheta$. $P(d)$ is a normalization constant, called the marginal likelihood: $P(d) = \int p(d|\btheta) \, p(\btheta) \, d\btheta$. Under the assumption of the data mentioned above, the likelihood function $P(d \, | \, {\btheta})$ can be written as:
\begin{equation}
P (d \, | \, {\btheta}) \propto \exp \left[ -\frac{1}{2} \langle d - h({\btheta}), d - h({\btheta}) \rangle \right],
\end{equation}
where $\langle a, b \rangle$ describes the noise-weighted inner product defined as:
\begin{equation}
\langle a, b \rangle := 4 \Re \int_{f_\mathrm{low}}^{f_\mathrm{high}} \frac{\tilde{a}(f) \, \tilde{b}^*(f)}{S_n(f)}df
\end{equation}
where $\tilde{a}(f)$ denotes the Fourier transforms of $a(t)$, and the integration limits are defined by the sensitivity bandwidth of the detector, $f_\mathrm{low}$ and $f_\mathrm{high}$.

In this work, we consider a global 3-detector network of the two Advanced LIGO detectors at Hanford (H) and Livingston (L) and the Advanced Virgo detector (V) at Cascina, Italy. The Advanced LIGO detectors are assumed to be at a sensitivity described by their ``high-power, zero-detuning'' configuration~\cite{aLIGOZeroDetHighPower} whereas the Advanced Virgo detector PSD is assumed to the one described in [cite LIGO document LIGO-P1200087-v18]. Assuming that the noise between any two detectors is uncorrelated, the joint likelihood across the three detectors is written as a product of the likelihoods in each detector:
\begin{equation}
P (d \, | \, {\btheta}) = \prod_{I \epsilon {H,L,V}} P (d^{I} \, | \, {\btheta}).
\end{equation}

In this Bayesian framework, we proceed to estimate the posterior probability distribution of $\btheta$ by stochastically sampling over the parameter space, using a python-based affine-invariant ensemble sampler \textsc{emcee}~\cite{foreman2013emcee, goodman2010ensemble}. Subsequently, we marginalize over the \textit{nuisance} parameters to obtain the posterior distributions on the non-GR parameter set, ${\Delta \blambda}$ or $c_{\ell m}$. If the data is consistent with a binary black hole signal in GR, $P(\Delta \blambda \, | \, d)$ (or $P(\Delta c_{\ell m} \, | \, d)$) is expected to be consistent with zero. 

We assume uniform prior probability distributions on the chirp mass and mass ratio in the interval $M_c \in [1,200] M_\odot$ and $q \in [0.05,1.0]$. The prior on the location of the source is  assumed to be isotropic on the sphere of  the  sky,  with $P({d_L}) \propto d_{L}^{2}$ where $d_L \in [1,10000]$ Mpc. We use an  isotropic  prior  on  the  orientation  of  the  binary: $P({\iota,\varphi_0,\psi}) \propto \sin\iota$ with $\iota \in [0,\pi)$, $\varphi_0 \in [0,2\pi)$ and $\psi \in [0,\pi)$. For all other parameters in $\btheta$, we use uniform priors: $\alpha \in [0,2\pi)$, $\delta \in [0,2\pi)$ and $t_0 \in [-15,15]$. 

\section{Simulations of binary black holes in GR}
\label{sec:simulations}

\begin{figure*}[tbh]
	\begin{center}
 		\includegraphics[height=2.0in]{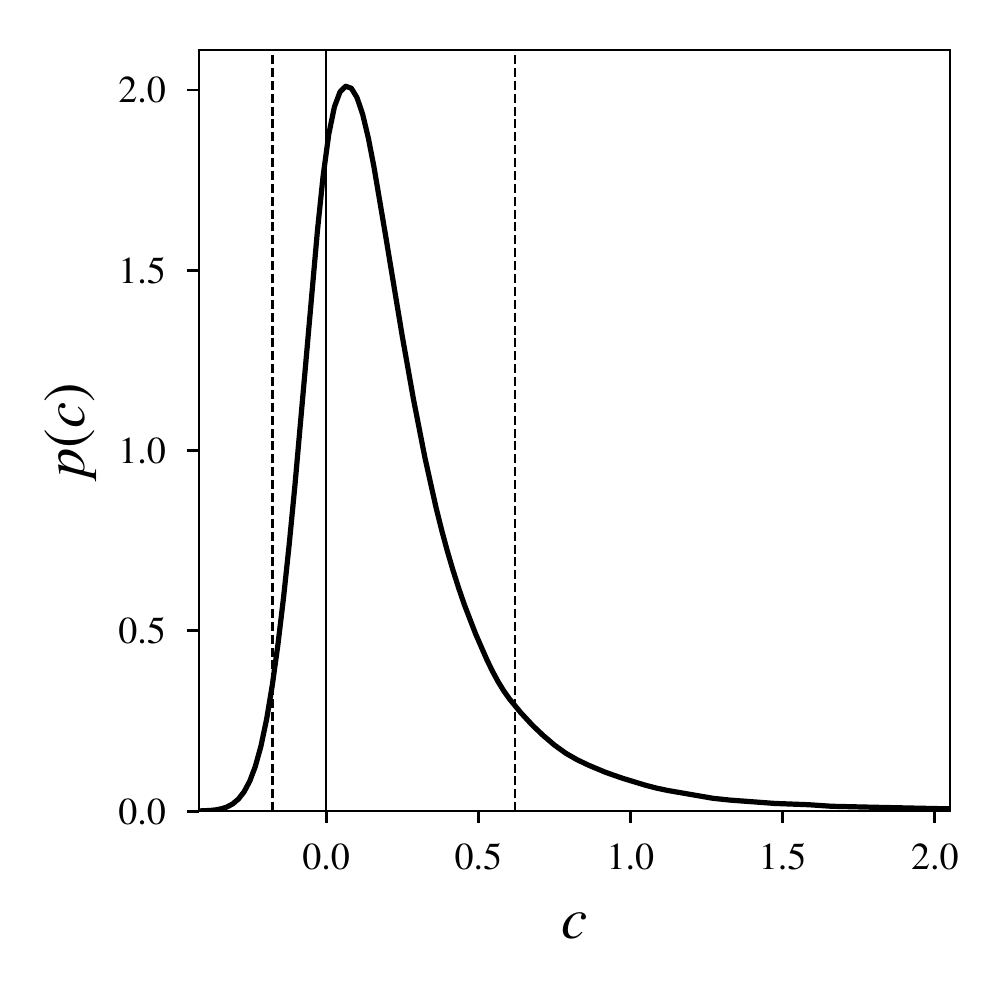}
		\includegraphics[height=2.45in]{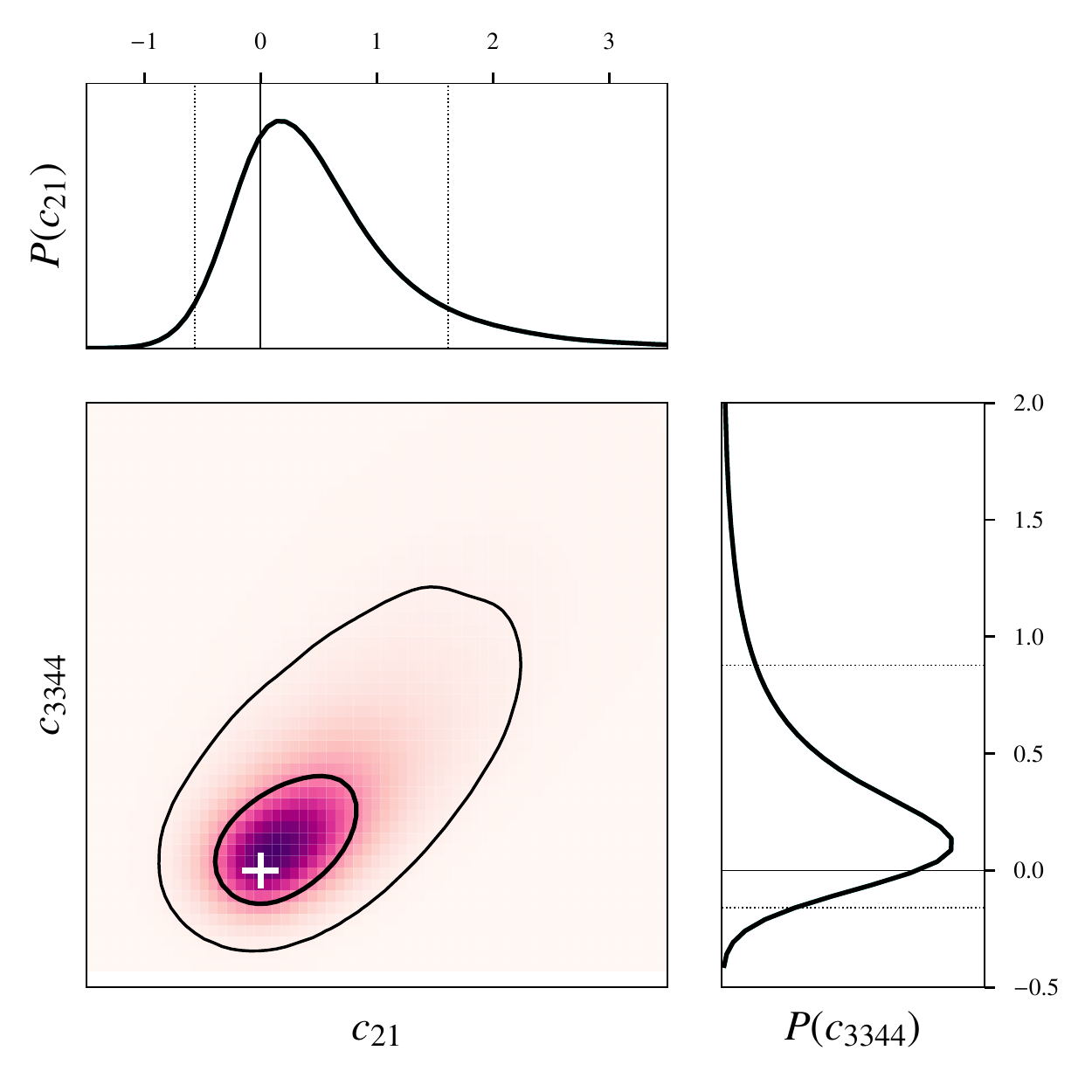}
		\includegraphics[height=2.45in]{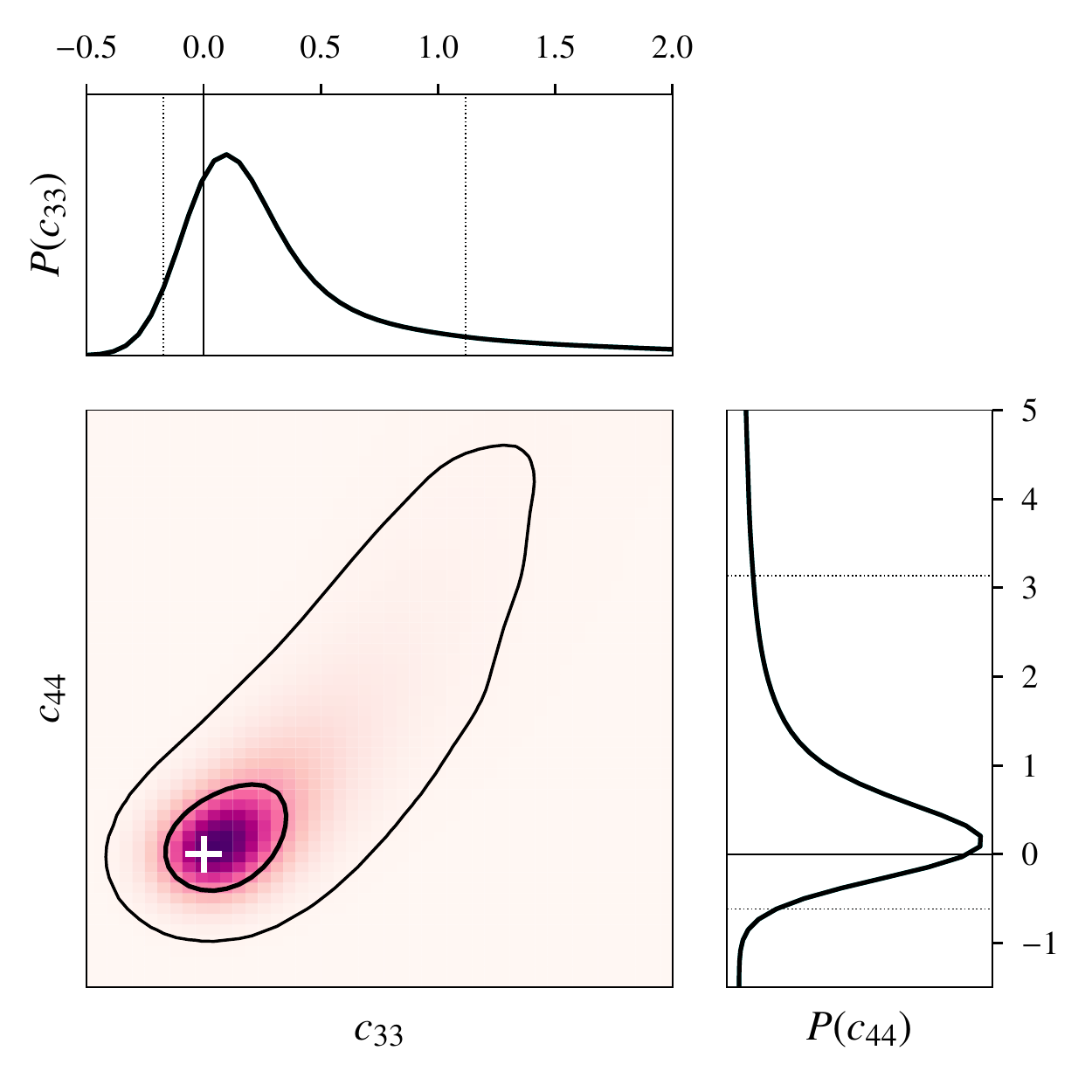}
	\end{center} 
 	\caption{\emph{Left:} The posterior probability distribution of the deviation parameter $c$ estimated from the same simulated GR observation in Fig.~\ref{fig:posterior_BBH_GR_inj} (version 1 of the test described in Sec.~\ref{sec:formulationB}). Thin black lines shows the expected value in GR. The dotted lines mark the 90\% credible regions. \emph{Middle:} Posteriors on $c_{21}$ and $c_{3344}$ from the same simulated observation (version 2 of the test). \emph{Right:} Posteriors on $c_{33}$ and $c_{44}$ from the same simulated observation (version 3 of the test).} 
	\label{fig:test2posts}
\end{figure*}

\begin{figure}[tbh]
	\begin{center}
		\includegraphics[width=2.8in]{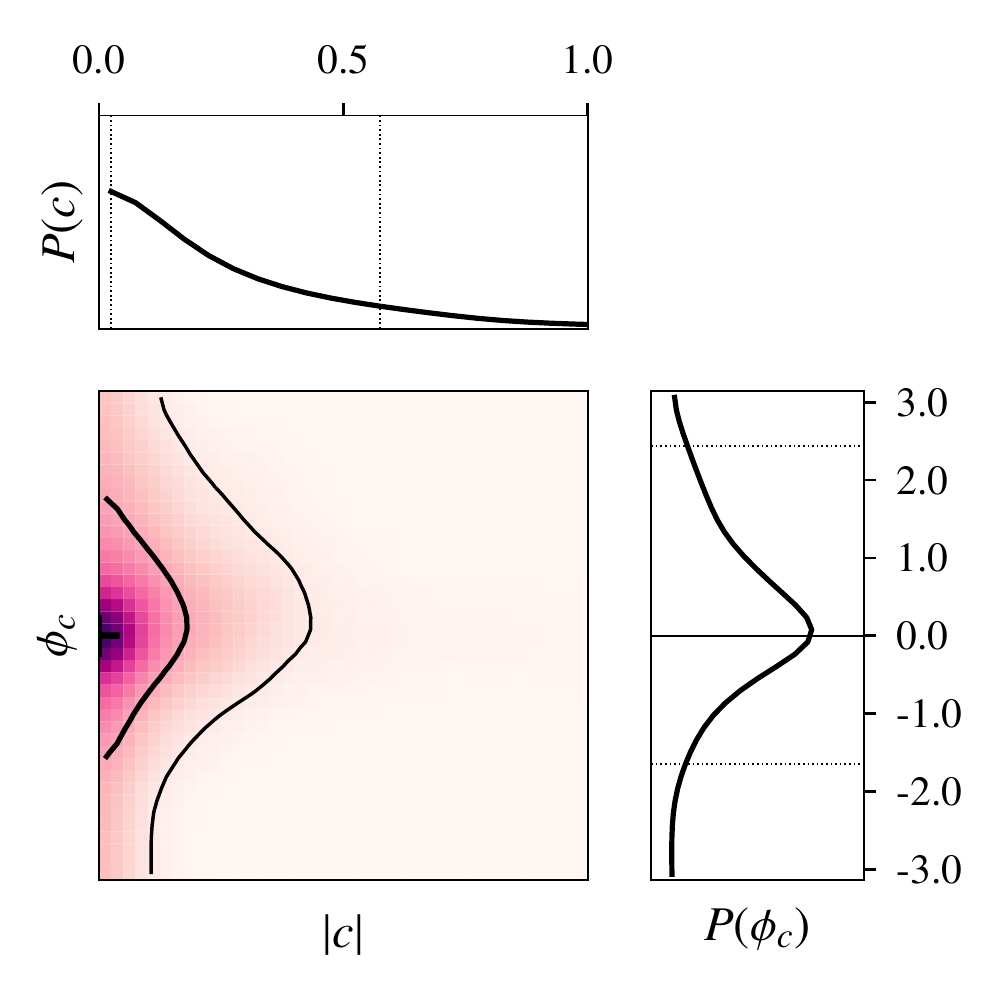}
	\end{center} 
	\caption{The figure shows the posterior probability distribution of the absolute value $|c|$ and argument $\phi_c$ complex deviation parameter $\tilde{c}$ estimated from the simulated GR event. Details are same as in \ref{fig:posterior_BBH_GR_inj}.}
	\label{fig:c1_complex}
\end{figure}

We use simulations of binary black hole events (as described in GR) to elaborate the two formulations of the tests presented in Sec.~\ref{sec:test}, i.e., by introducing extra parameters to describe the higher harmonics, and try to estimate and constrain them from the data, using a Bayesian framework.

\subsection{Formulation A:}
\label{sec:formulationA}
The first test we consider the formulation proposed in Eq.~\eqref{eq:test_1}. This follows the outline presented in~\cite{Dhanpal:2018ufk} to check for the consistency of intrinsic parameters $\blambda := \{M_c, q\}$ estimated from the dominant mode and the higher order modes. While ~\cite{Dhanpal:2018ufk} focuses on one performing this test with only one Advanced LIGO detector, we study the performance of this test in the case of the three detector Advanced LIGO-Virgo network. 

We consider two different ways to perform the test. First, we introduce \emph{one} deviation parameter at a time. That is, $\Delta\blambda = {\Delta M_c}$ or $\Delta\blambda = {\Delta q}$. We then consider introducing a concurrent deviation in \emph{two} parameters $\Delta \blambda = \{\Delta M_c, \Delta q\}$. In Fig.~\ref{fig:posterior_BBH_GR_inj}, we show the results of the tests performed with GR waveform by varying either one parameter or two parameters, for a binary with total mass $M = 80M_{\odot}$, mass ratio $q=1/9$, inclination angle $ {\iota}=60^{\circ} $ producing a network signal-to-noise ratio  (SNR)  of 25 (SNR in higher modes is $\sim 10$). SNRs in individual detectors are: 15 in Advanced LIGO-Hanford, 18.9 in Advanced LIGO-Livingston and 6.7 in Virgo. The posterior probability density for both the parameters $\Delta q$ and $\Delta M_c$ are consistent with zero as one expect in GR. Furthermore, the deviation parameters are found to be better constrained when only one deviation parameter is allowed to vary at a time (either $\Delta M_c$ or $\Delta q$). This suggests that a consistency test with only one deviation parameter in the higher modes would provide tighter constraints on deviations. In the subsequent analysis, we therefore focus on varying only one deviation parameter at a time. 

In Fig.~\ref{fig:hm_mcq_compare-1det_3det_GR_inj} we show that, as expected, the width of the posteriors of the deviation parameters become smaller (i.e., improved precision) when we perform the test with a network of three Advanced LIGO-Virgo detectors instead of using only one Advanced LIGO detector (for the same SNR). However, for a fixed SNR, the improvements in the precision is small (factor of $\sim ~2-3$), due to the fact that the improved information (e.g., sky localization) is not highly correlated with the intrinsic parameters $M_c, q$ nor the deviation parameters $\Delta M_c, q$. 

Figures~\ref{fig:delmc_delq_varyingM} and \ref{fig:delmc_delq_varyingq} show the 90\% credible intervals of the posteriors of the deviation parameters for binaries with varying masses, mass ratios and inclination angles, estimated using the three detector network. In all cases, we set the network SNR to be {25}. Note that only one deviation parameter ($\Delta M_c$ or $\Delta q$) is varied at a time.  We find that binaries with large mass ratios ($q < 1/ 2$) and inclination angles ($\iota > 60 ^\circ $) will allow precision tests of the GR predictions, reaching statistical uncertainties of $< 10^{-3}$ for $\Delta q$ and $< 10^{-2}$ for the dimensionless deviation parameter $\Delta M_c/M_c$. Our results are found to be consistent with the one detector analysis done in ~\cite{Dhanpal:2018ufk}. We, however, notice that the 90\% interval for both the deviation parameters, in three detector analysis, decreases slightly (i.e., precision improved) as compared to the one detector case.
 
\subsection{Formulation B}
\label{sec:formulationB}

\begin{figure}[h]
	\begin{center}
		\includegraphics[scale=0.85]{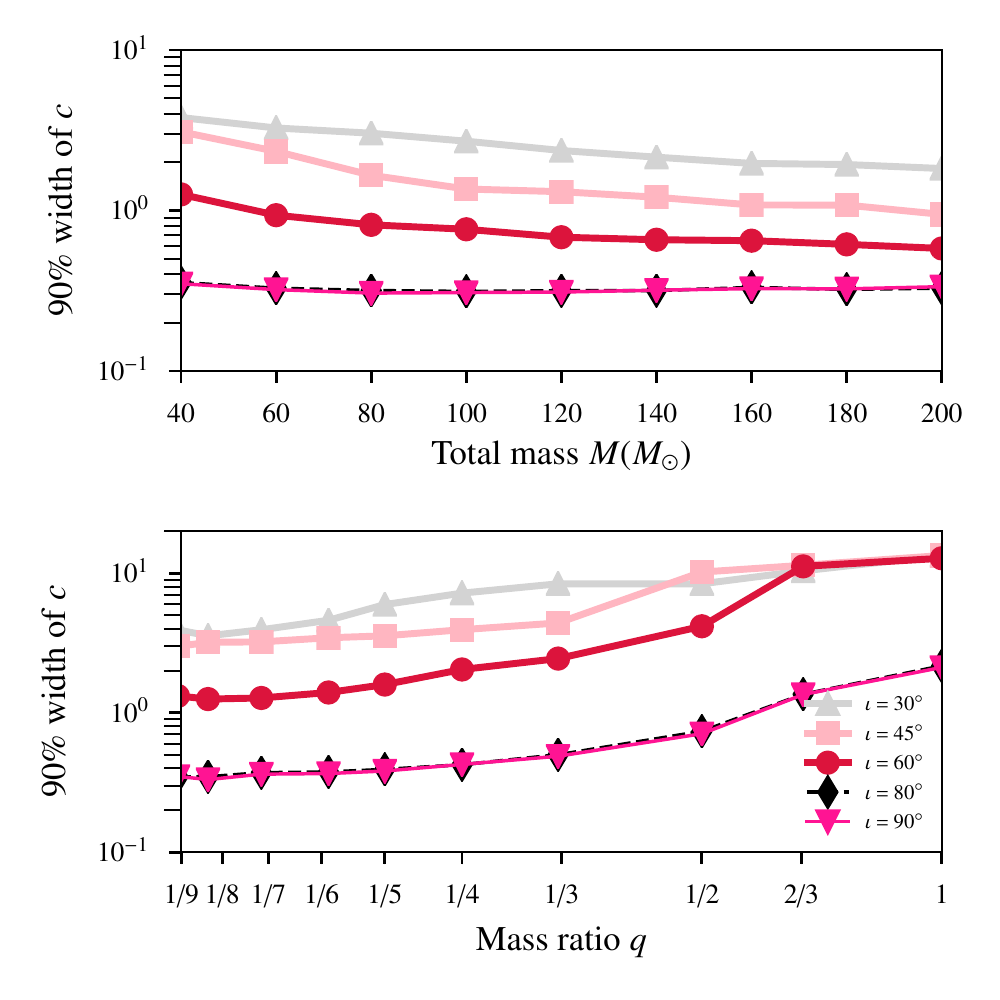}
	\end{center} 
	\caption{The width of 90\% credible regions of the posteriors of $c$ for binaries with different total mass $M$ (upper panel) and mass ratio $q$ (lower panel) and inclination angle $\iota$ (legends). All binaries considered in the upper panel have a mass ratio $q=1/9$. Binaries considered in the lower panel have total mass of $40M_{\odot}$. All the simulated observations produce a network SNR of 25 in Advanced LIGO-Virgo network.}
	\label{fig:constraint_c}
\end{figure}
 
\begin{figure*}[tbh]
	\begin{center}
		\includegraphics[scale=0.75]{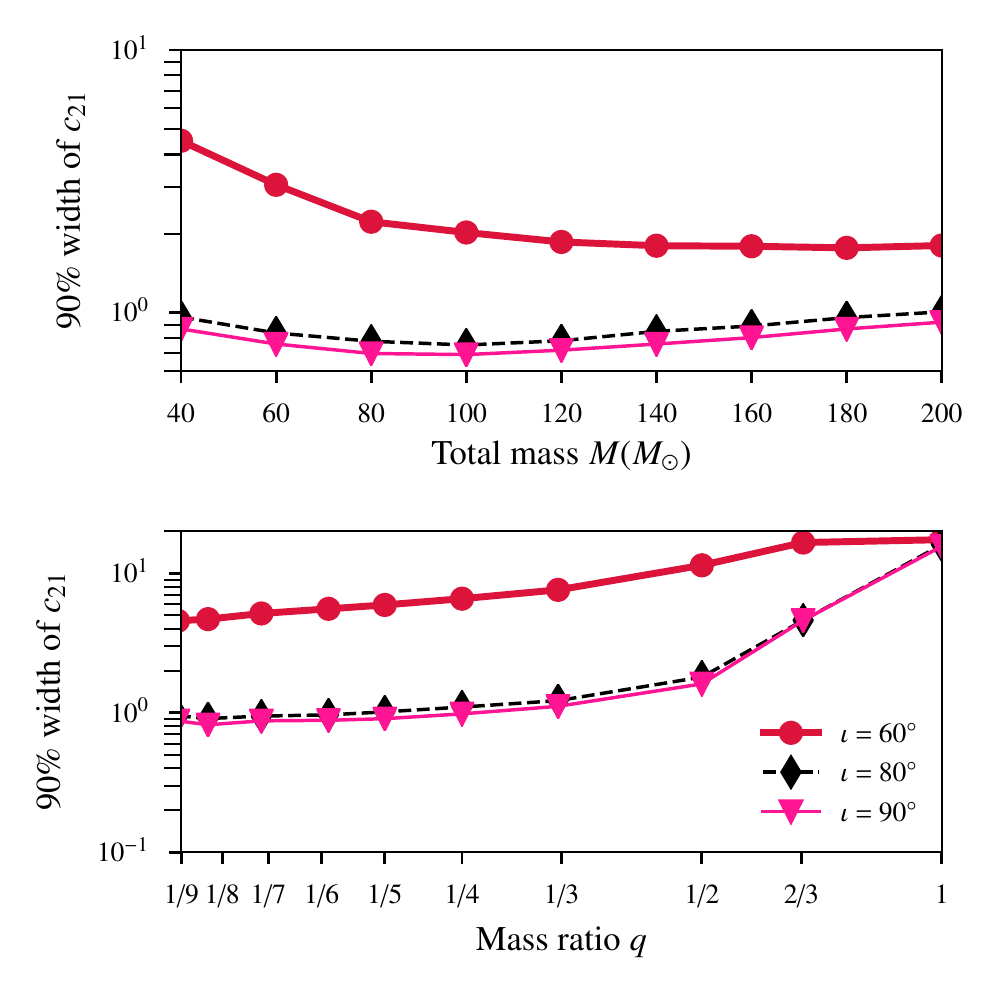}
		\includegraphics[scale=0.75]{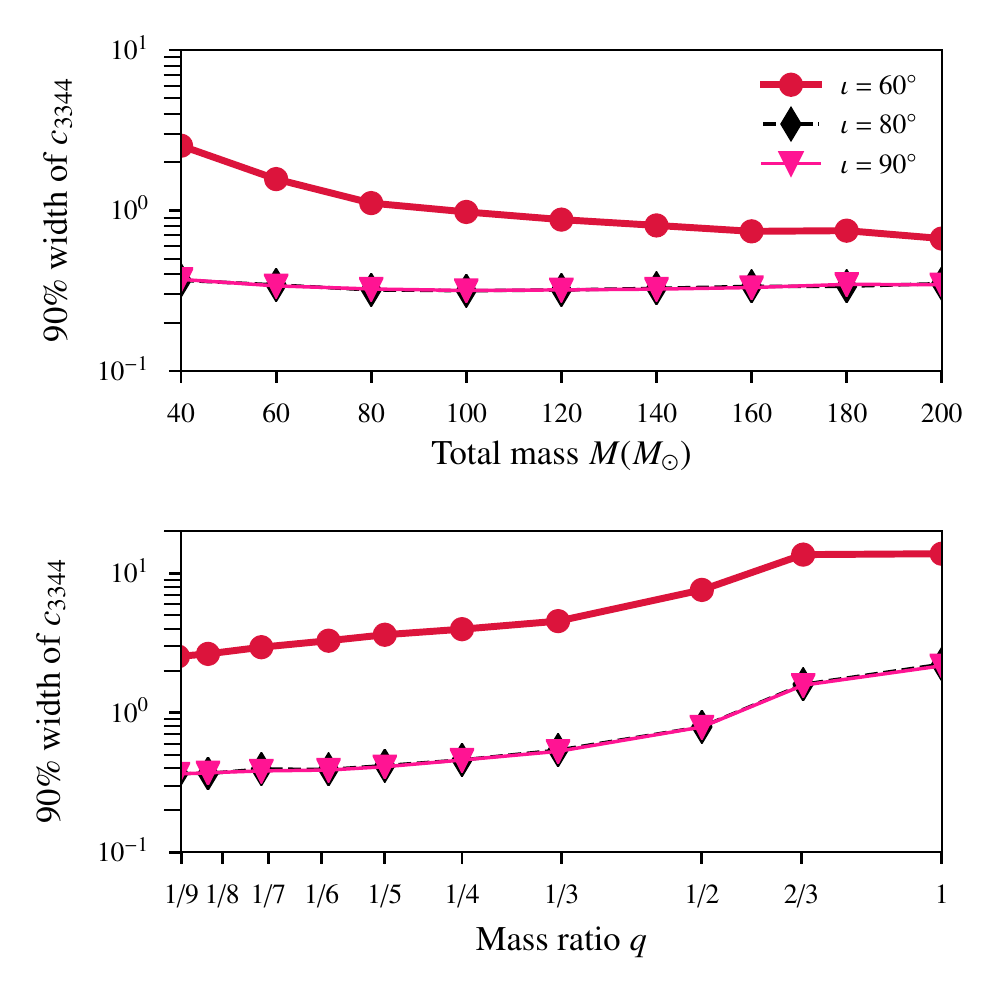}
	\end{center}
	\caption{Same as Fig.~\ref{fig:constraint_c} except that the posteriors are of the deviation parameters $c_{21}$ (left plots) and  $c_{3344}$ (right plots).}
	\label{fig:constraint_c21_c3344}
\end{figure*}

\begin{figure*}[tbh]
	\centering
	\includegraphics[scale=0.75]{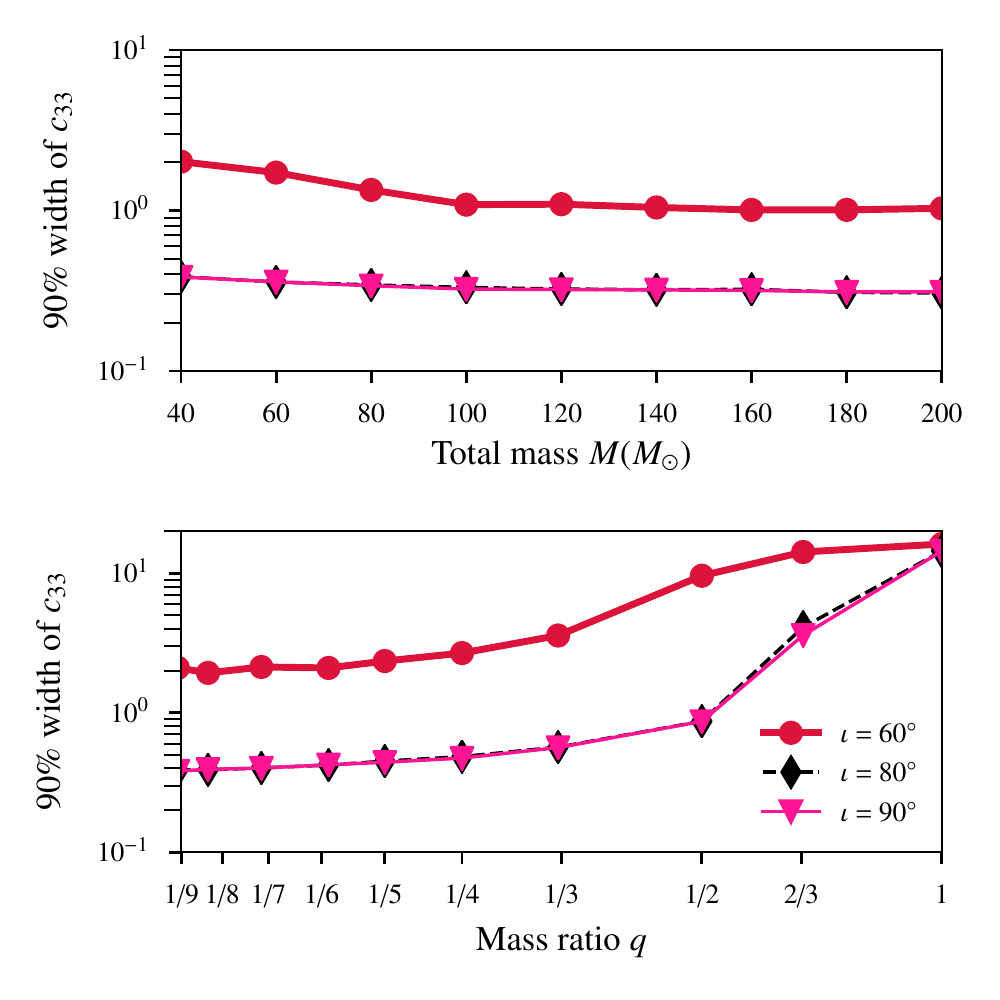}
	\includegraphics[scale=0.75]{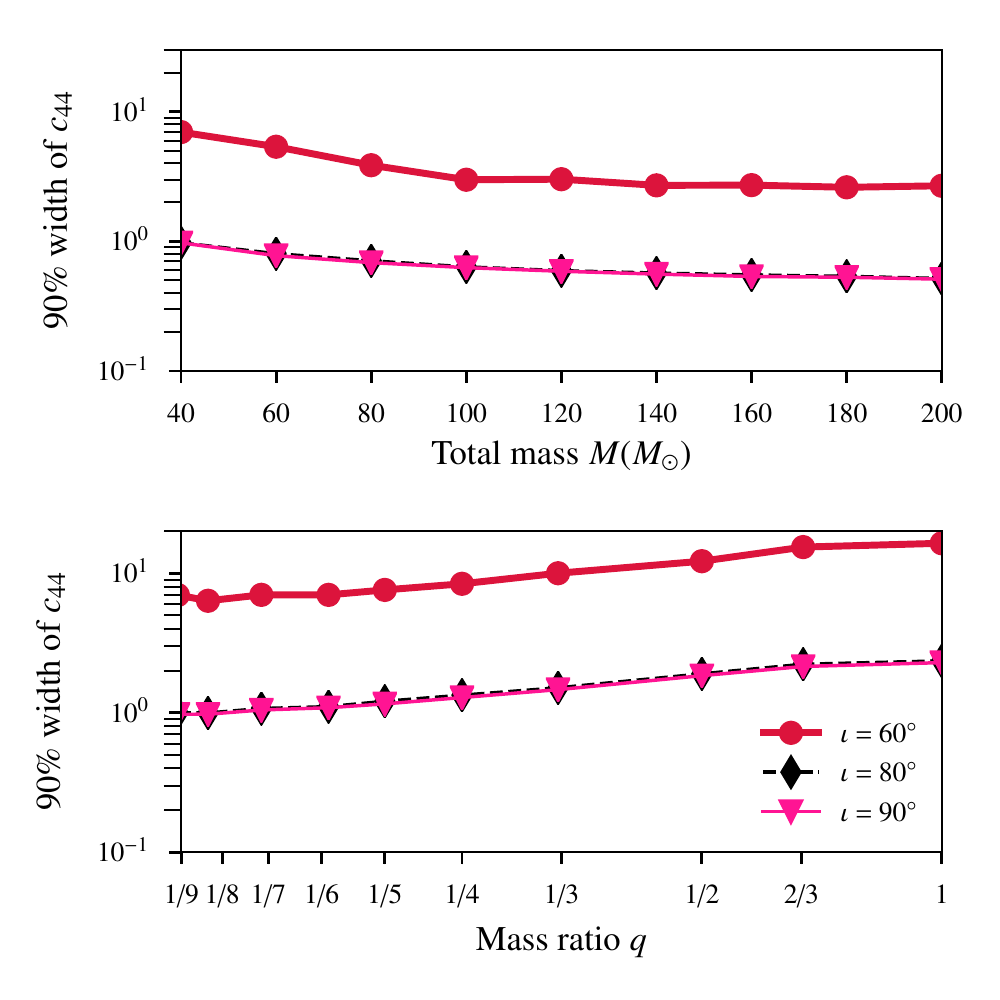}
	\caption{Same as Fig.~\ref{fig:constraint_c} except that the posteriors are of the deviation parameters $c_{33}$ (left plots) and  $c_{44}$ (right plots).}  
	\label{fig:constraint_c33_c44}
\end{figure*}

Now we consider the formulation proposed in Eq.~\eqref{eq:test_2}, which involves introducing generic possible deviation parameters $c_{\ell m}$ in the amplitudes of the higher order modes. Indeed, the most general form of this test would treat all the $c_{\ell m}$ as free parameters. However, because of the correlation among these parameters and with some of the other parameters of the binary (such as the luminosity distance and inclination angle), this is likely to result in poor constraints on these parameters. Hence we consider different flavors of this test. 

\begin{enumerate}
\item We set $c := c_{21} = c_{33} = c_{44}$ and estimate the posteriors of $c$ along with all other binary parameters present in the GR waveform.  
\item We allow $c_{21}$ and $c_{3344} := c_{33} = c_{44} $ to vary and estimate the posteriors of $c_{21}$ and $c_{3344}$ along with all other binary parameters present in the GR waveform.  
\item We fix $c_{21} = 0$ and vary $c_{33}$ and $c_{44}$, thus estimating the posteriors of $c_{33}$ and $c_{44}$. 
\end{enumerate}

In Fig.~\ref{fig:test2posts}, we show example posteriors of the deviation parameters obtained from a simulated binary black hole system (in GR) with  a total mass $M = 80M_{\odot}$, mass ratio $q=1/9$ and inclination angle $ {\iota}=60^{\circ} $, producing an SNR of 25 in the Advanced LIGO-Virgo network. The left plot shows the posterior of the deviation parameter $c$ (version 1 of the test), while the middle plot show the posteriors of $c_{21}$ and $c_{3344}$ (version 2 of the test) and the right panel shows the posteriors on $c_{33}$ and $c_{44}$  (version 3 of the test). We see that all the posterior distributions are consistent with zero. 

A more general version of these tests with amplitude correction in the higher modes would be to assume that the deviation parameters are complex in nature i.e. they have a magnitude as well as a phase component. To demonstrate such test, we replace the real amplitude correction $c$ (version 1 of the test) with a complex correction $\tilde{c} = |\tilde{c}| \, e^{\phi_c}$. Figure~\ref{fig:c1_complex} shows the posterior probability distribution of both the magnitude and phase of the deviation parameter $\tilde{c}$ from the same simulated GR event described in Fig.~\ref{fig:test2posts}. We find that though the absolute value of complex correction is well constrained, the phase remains uninformative. Hence for all the future tests we restrict to real valued deviation parameters. 

Figure \ref{fig:constraint_c} shows the width of the 90\% credible regions in the posterior of $c$  (version 1 of the test) as a function of the total mass and mass ratio of the binary (producing network SNR of 25 in all cases). Figure~\ref{fig:constraint_c21_c3344} shows the width of the 90\% credible regions in the posteriors of $c_{21}$ and  $c_{3344}$  (version 2 of the test) while Fig.~\ref{fig:constraint_c33_c44} shows the same for $c_{33}$ and  $c_{44}$  (version 3 of the test).

We observe that the constraints on the deviation parameters become narrower for binaries with larger mass ratios and inclination angles. We find that $c$ is, in general, better constrained than $\{c_{21}, c_{3344}\}$ and $\{c_{33}, c_{44}\}$. However, the statistical uncertainties in $c$, $\{c_{21}, c_{3344}\}$ and $\{c_{33}, c_{44}\}$ are modest, reaching only $\sim$ 1 (as opposed to the parameters discussed in Sec.~\ref{sec:formulationA}, which can be constrained to a precision of $\sim 10^{-2}$). The statistical precision of these tests largely depends on the signal-to-noise distribution in the higher modes. These constraints could be significantly improved with third-generation ground based detectors or space based detectors as they will detect hundreds of signals with good SNR and, in turn, enhance the precision of parameter estimation. The low SNR in the higher modes has resulted in posteriors wider than the priors for $\{c_{21}, c_{3344}\}$ and $\{c_{33}, c_{44}\}$ for $\iota={30^{\circ},45^{\circ}}$. Hence these results are not presented.

\section{Simulations with deviations from binary black holes in GR}
\label{sec:simulation_nonbbh}

\begin{figure}[htb] 
\begin{center}
\includegraphics[width=3.4in]{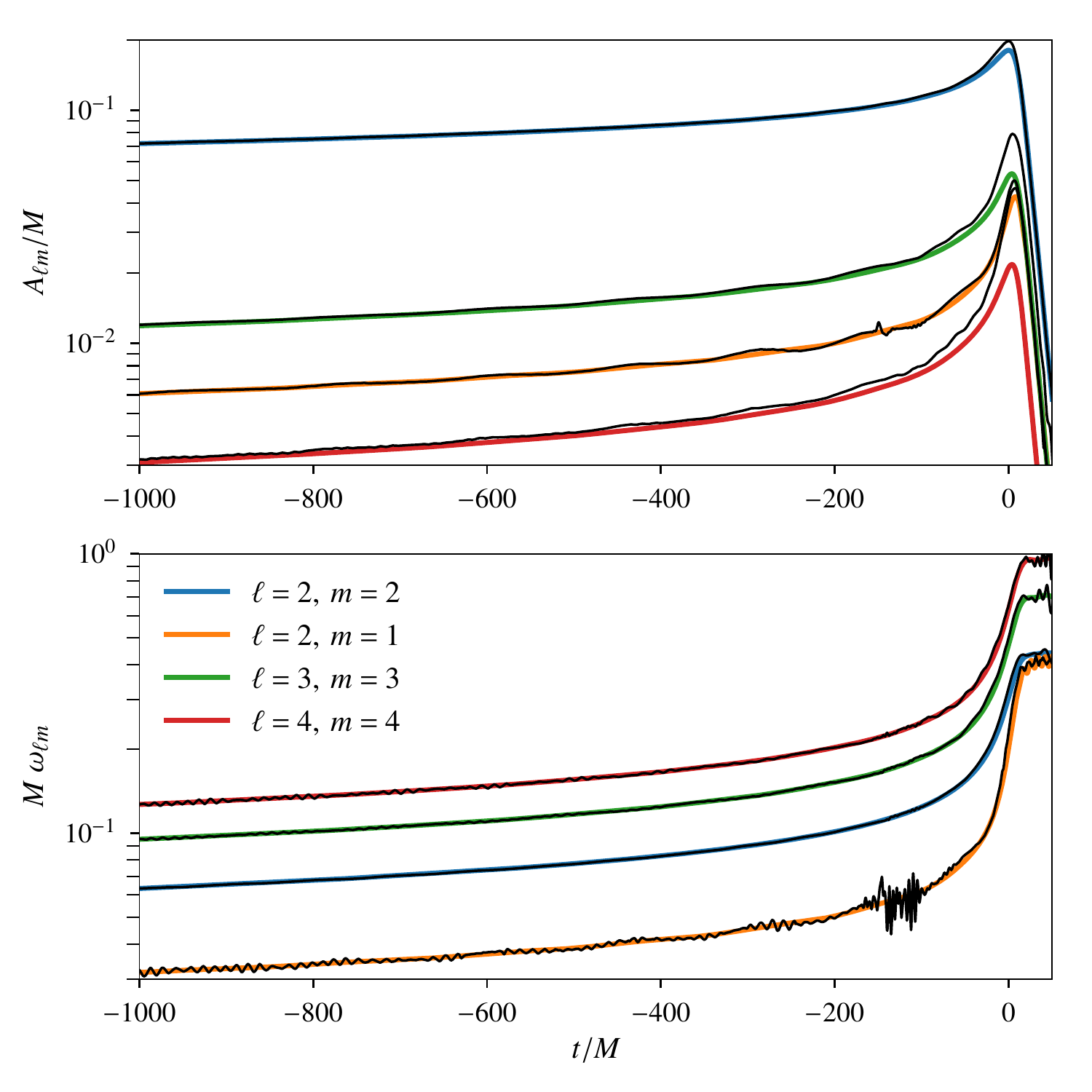}
\end{center} 
\caption{Colored traces show the time domain amplitude $A_{\ell m} := |\hlm|$ (top panel) and instantaneous angular frequency $\omega_{\ell m} := d\phi_{\ell m}/dt$ of different modes (shown in legend) of a non-spinning binary black hole waveform with mass ratio $q = 1/6$. The black traces show the same for a re-scaled numerical relativity waveform from a neutron star-black hole simulation. The small oscillations here are numerical artifacts in the simulated waveform.}
\label{fig:bbh_nsbh_waveforms}
\end{figure}

\begin{figure}[htb] 
\begin{center}
\includegraphics[width=3.4in]{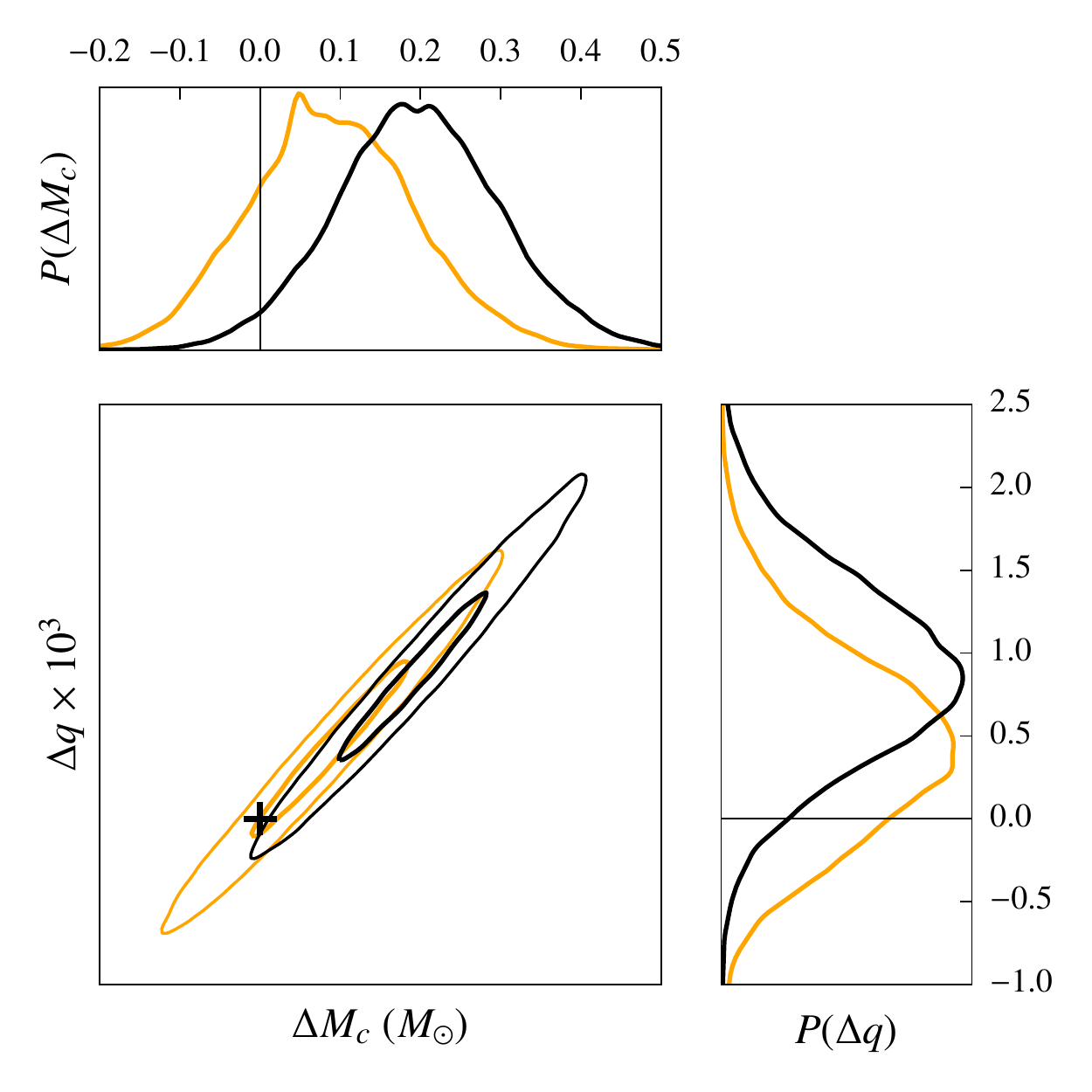}
\end{center} 
\caption{Posterior distributions of $\Delta M_c$ and $\Delta q$ estimated from simulated GW observation of a non-binary black hole system (black contours) with $M = 120 M_\odot, q = 6$ (obtained by re-scaling the NS-BH waveform SXS:BHNS:0001 from the SXS catalog) and a black hole system with the same parameters (yellow contours). The black `+' sign in the middle panel, the black vertical line in the top panel, and the black horizontal line in the right panel, indicate the expected value of $\Delta M_c = 0, \Delta q = 0$ for a binary black hole system in GR.} 
\label{fig:non_bbh_posterior}
\end{figure}

In this section we demonstrate that, if the multipole structure of the radiation is sufficiently different from that of a binary black hole system in GR (either when the underlying theory is different from GR or when the binary contains compact objects other than black holes), then this test should be able to identity this. Note however, that this will require the higher order modes to be observed with sufficient SNR, which typically happen for the case of massive binaries with large mass ratio observed with large inclination angles. Thus, this test is unlikely to distinguish black hole - neutron star binaries from binary black holes as the total mass of the system is unlikely to be greater than $\sim 50 M_\odot$ (going by the mass distribution of the black holes observed by LIGO and Virgo so far). Hence we rescale the gravitational waveform produced by the numerical relativity simulation of a non-spinning black hole-neutron star binary to a larger total mass so that the higher modes are observed with sufficient SNR. We use the  black hole-neutron star waveform with mass ratio ${1/6}$ from the numerical-relativity waveform catalog of the SXS collaboration~\cite{SXS-Catalog} (SXS:BHNS:0001; with component masses $8.4 M_\odot$ and $1.4 M_\odot$). We rescale this waveform to a total mass of $M = 120 M_\odot$ and use it as a proxy for a gravitational waveform from a binary consisting of at least one non-black hole compact object. Note that the rescaled signal will not correspond to a black hole-neutron star binary, as $m_2 \simeq 17 M_\odot$ is much larger than the maximum mass of a neutron star.  Figure~\ref{fig:bbh_nsbh_waveforms} compares the amplitude $|\hlm|(t)$ and instantaneous frequency $d\phi_{\ell m}(t)/dt$ of this waveform, along with a similar waveform from a non-spinning binary black hole system with the same mass ratio. The multipole structure of these waveforms can be seen to be slightly different. We hope that the test will be able to identify these differences provided higher modes are observed with sufficient SNR. 

Figure~\ref{fig:non_bbh_posterior} shows the posteriors of the deviation parameters $\Delta M_c$ and $\Delta q$ estimated from a simulated observation containing this signal (darker contours), which are {inconsistent} with the GR prediction of binary black holes ($\Delta M_c = \Delta q = 0$). The Figure also shows the results of the test applied on a numerical relativity waveform from a binary black hole system with same parameters (lighter contours), which shows consistency with $\Delta M_c = \Delta q = 0$. The simulated signals correspond to binaries with inclination angle $\iota = 90^\circ$, producing SNR of 50 in the three detector Advanced LIGO-Virgo network.

\section{Waveform systematics}
\label{sec:waveformsyst}

In all the simulations presented in the previous section, we have assumed that binary black holes have negligible spin angular momenta. While most of the binary black hole events detected by LIGO and Virgo do not appear to have significant spins~\cite{abbott2019gwtc}, black holes in binaries, in general, could be spinning. When non-spinning waveform templates are employed to perform the consistency test on GW observations of spinning binaries, the incomplete modeling of the templates can manifest as a deviation from the predicted behavior of a binary black hole signal in GR. Here we make a first estimate of the effect of neglecting black hole spins in this test by performing the same analysis on simulated spinning binary black hole observations. We simulate spinning binary black hole observations making use of the numerical-relativity surrogate waveform family developed in~\cite{varma2019surrogate} and perform the consistency test using the same non-spinning waveform family~\cite{Mehta:2017jpq} as the base GR waveform over which modifications are applied. We focus on the Formulation A (see Sec.~\ref{sec:formulation}) as this formulation yields the tightest constraints on deviations from the predicted behavior and hence is most prone to systematic errors. 

\begin{figure*}[tbh] 
\begin{center}
\includegraphics[width=3.3in]{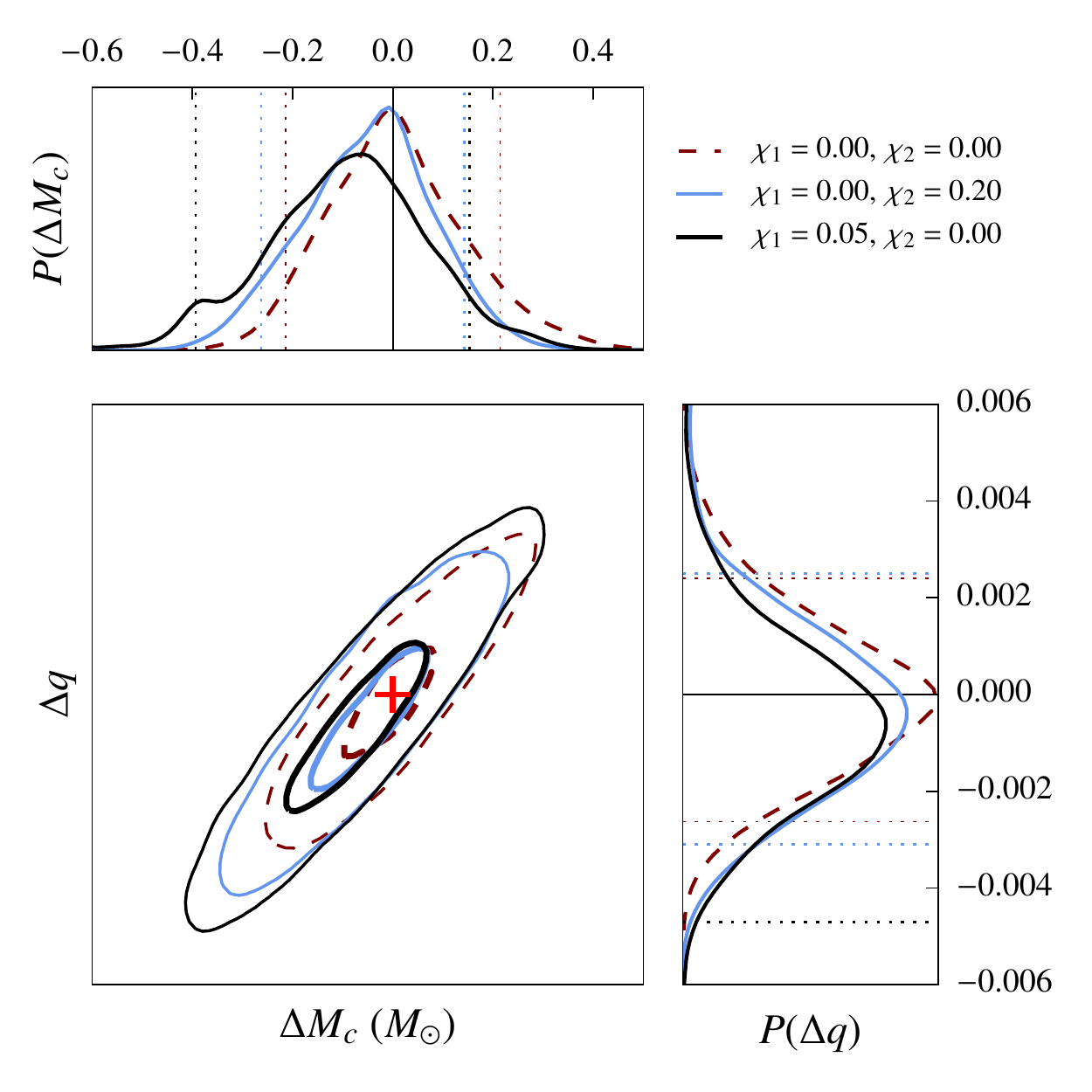}
\includegraphics[width=3.3in]{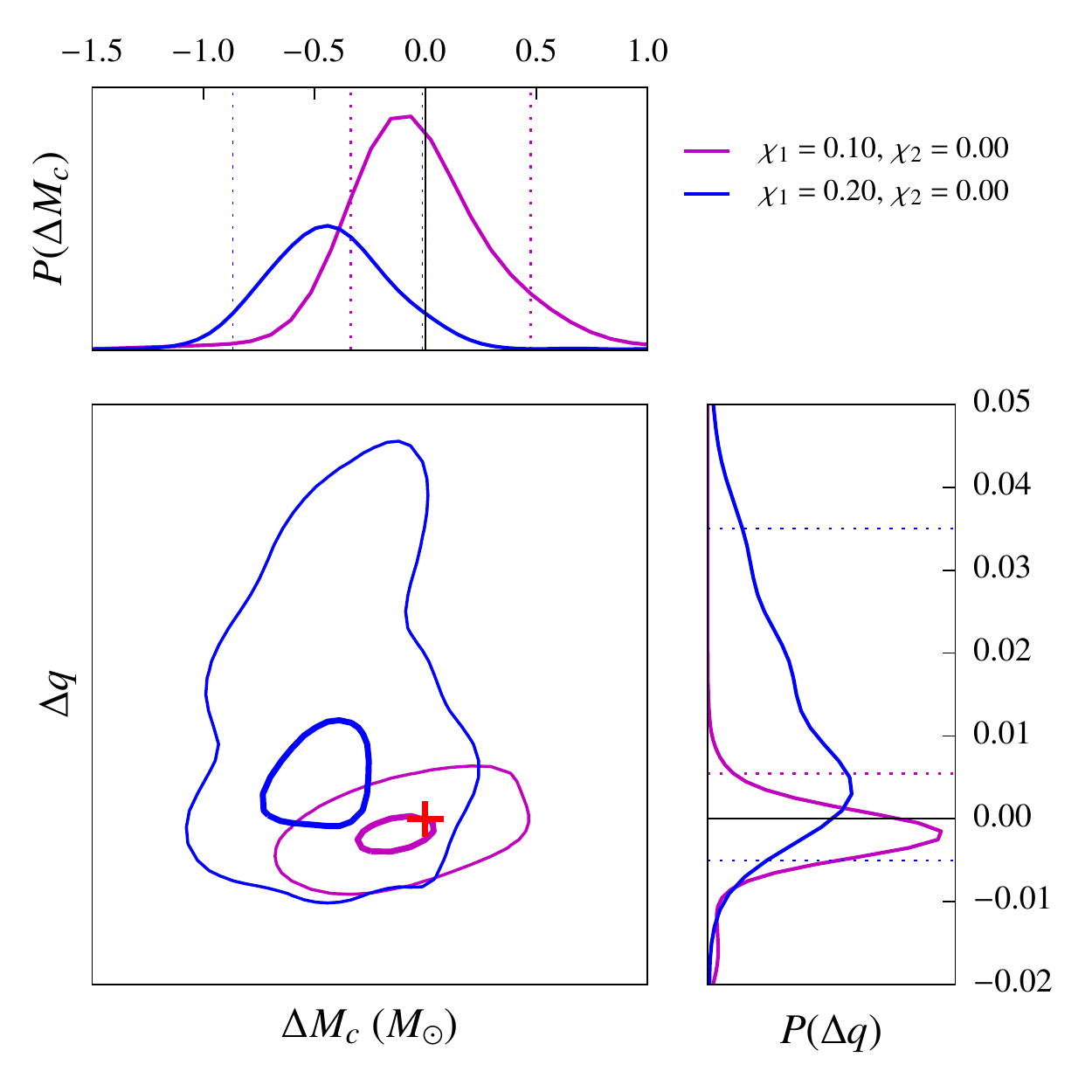}
\end{center} 
\caption{Posteriors of the deviation parameters $\Delta M_c$ and $\Delta q$ for binary black hole injections with different values of component spins $\chi_1$ and $\chi_2$ (shown in legends). The left plot corresponds to low spins and the right to high spins. The results from non-spinning injections is also shown, for comparison (left). We see that width of the posteriors from the highly spinning injections are much large as compared to that from non-spinning or low-spin injections (note the different axis ranges in the left and right plots).} 
\label{fig:posterior_spin}
\end{figure*}

Figure~\ref{fig:posterior_spin} shows the posteriors in the deviation parameters $\Delta M_c$ and $\Delta q$ introduced in Eq.\eqref{eq:test_1} estimated from simulated binary black hole events with different values of spin (dimensionless spins $\chi_{1,2}$ shown in legends). The left plot corresponds to simulations with low spins, while the right plot to high spins.  For high spin injections, though the posteriors of $\Delta M_c/M_c$ and $\Delta q$ broadly are consistent with GR value (0, 0) at the 90\% level, the peaks of the posteriors show a bias from the injected value. Additionally, the widths of the deviation parameters increase significantly for high spins of the primary black hole. This suggests that one should use an accurate spinning waveform model if one wants to perform such tests on highly spinning signals. 

\section{Discussions \& Conclusion}
\label{sec:conclusions}

In this paper, we have proposed a set of tests of the ``no-hair'' nature of binary black holes in GR based on a consistency test of the multipolar structure of the gravitational radiation. These tests are analogous to the tests of ``no-hair'' theorem for stationary black holes based on the consistency of different quasi-normal modes of a perturbed black hole~\cite{xx}. We proposed two formulations of this test, that introduce extra deviation parameters that govern the amplitude and phase evolution of different spherical harmonic modes of the radiation, as well as ones affecting the amplitudes of different modes. Posterior distributions of these deviation parameters can be estimated using a Bayesian framework. 

The first formulation is inspired by the fact that different modes of radiation from the binary black holes should be uniquely described only by the same values of intrinsic parameters (chirp mass and mass ratio), and hence these parameters estimated from different modes should be consistent to each other. We first revisited this formulation, originally presented in~\cite{Dhanpal:2018ufk}. We presented the results expected from 3-detector observations of binary black holes using the Advanced LIGO-Virgo detectors. Results from our simulations suggest that upcoming observations using Advanced LIGO and Virgo will be able to put precise constraints on the deviation parameters. Indeed, this test requires appreciable SNR in the higher order modes of the observed GW signal, which is expected only for small fraction (a few percents~\cite{Dhanpal:2018ufk}) of detectable binary black hole events. However, given that LIGO-Virgo would observed hundreds of binary black hole mergers in coming years, we expect a reasonable number of such events to be observed. We also demonstrate that, if the observed signal is not produced by a binary black hole system in GR, the test is able to identify this, provided that the SNR is high enough.

In the second formulation, we check for the consistency between the amplitudes of different modes. In order to do so, we introduce a set of extra deviation parameters in the amplitudes for the higher modes. We see that these deviation parameters can be constrained only with modest precision in Advanced LIGO-Virgo. However, the precision of such a test is expected to increase manifold with the next generation of detectors (e.g. with Einstein Telescope or LISA). 

We also presented a preliminary investigation of the effect of neglecting the effect of black hole spin in the analysis and find that if the binary has significant effective spin, neglecting spin effects can produce a bias in the estimated posteriors. This can mimic a deviation from the no-hair nature of binary black holes. Thus, applying this test to real GW data will require the use of accurate waveform templates that include non-quadrupole modes and spin effects, which are starting to become available now~\cite{london2018first,cotesta2018enriching}. 

Note that all of the binary black hole detections during the first two observing runs of the Advanced LIGO-Virgo network have been consistent with equal or almost equal-mass systems with inclinations that are close to face-on/face-off. Thus, they are not expected to have sufficient contribution from higher modes to perform the test proposed in this paper. However, with increasing sensitivity of the current generation of detectors in the coming years, we expect to detect GW signals from binary black holes which are highly asymmetric and/or highly inclined. From them we expect this test to give significant constraints on deviations from the predicted multipolar structure. 

\bigskip 
\acknowledgments
We thank Harald Pfeiffer, Bala Iyer, K. G. Arun and Gregorio Carullo for useful discussions, and Chandra Kant Mishra for help with the numerical implementation of the waveform model used in this paper. This research was supported by the Indo-US Centre for the Exploration of Extreme Gravity funded by the Indo-US Science and Technology Forum (IUSSTF/JC-029/2016). PA's research was, in addition, supported by the Max Planck Society through a Max Planck Partner Group at ICTS-TIFR and by the Canadian Institute for Advanced Research through the CIFAR Azrieli Global Scholars program. BSS's research was supported by NSF Grants AST-1716394 and AST-1708146. Computations were performed at the ICTS cluster Alice. 

\bibliography{TestGR.bib}

\end{document}